\documentclass[preprint]{imsart}

\usepackage{amsthm,amsmath,natbib}
\RequirePackage[colorlinks,citecolor=blue,urlcolor=blue]{hyperref}
\usepackage{epsfig}
\usepackage[utf8]{inputenc}
\usepackage{amssymb}
\usepackage{amsfonts}
\usepackage{amsthm}
\usepackage{amsmath,multirow}
\usepackage[a4paper]{geometry}
\usepackage{bm}
\usepackage{color}
\usepackage{paralist}

\startlocaldefs

\newtheorem{theorem}{Theorem}[section]

\theoremstyle{definition}
\newtheorem{definition}[theorem]{Definition}

\theoremstyle{remark}

\numberwithin{equation}{section} \numberwithin{theorem}{section}

\renewcommand{\mathbf}{\boldsymbol}

\newcommand{\set}[1]{\left\{#1\right\}}

\newcommand{\R}{\mathbb R}

\newcommand{\N}{\mathbb N}
\newcommand{\Z}{\mathbb Z}

\newcommand{\eps}{\varepsilon}

\newcommand{\BM}{{\scriptscriptstyle\mathrm{BM}}}
\newcommand{\POT}{{\scriptscriptstyle\mathrm{POT}}}
\newcommand{\PWM}{{\scriptscriptstyle\mathrm{PWM}}}
\newcommand{\ML}{{\scriptscriptstyle\mathrm{ML}}}
\newcommand{\GEV}{{\scriptscriptstyle\mathrm{GEV}}}

\endlocaldefs

\begin{document}

\begin{frontmatter}

\title{A horse racing between the block maxima method and the peak--over--threshold approach}
\runtitle{Block Maxima \textit{vs.} Peak-over-Threshold}

\author{\fnms{Axel} \snm{B\"{u}cher}\ead[label=e1]{axel.buecher@hhu.de}}
\address{Heinrich-Heine-Universit\"at D\"usseldorf,
Mathematisches Institut,
Universit\"atsstr.~1, 40225 D\"usseldorf, Germany. \printead{e1}}
\affiliation{Heinrich-Heine-Universit\"at D\"usseldorf}
\and
\author{\fnms{Chen} \snm{Zhou}\ead[label=e2]{zhou@ese.eur.nl; c.zhou@dnb.nl}}
\address{Econometric Institute,
Erasmus University Rotterdam,
3000 DR Rotterdam, The Netherlands. \printead{e2}}
\affiliation{Erasmus University Rotterdam and De Nederlandsche Bank}

\runauthor{B\"ucher and Zhou}

\begin{abstract}
Classical extreme value statistics consists of two fundamental approaches: the block maxima (BM) method and the peak-over-threshold (POT) approach. It seems to be general consensus among researchers in the field that the POT method makes use of extreme observations more efficiently than the BM method.
We shed light on this discussion from three different perspectives. First, based on recent theoretical results for the BM approach, we provide a theoretical comparison in i.i.d.\ scenarios. We argue that the data generating process may favour either one or the other approach. Second, if the underlying data possesses serial dependence, we argue that the choice of a method should be primarily guided by the ultimate statistical interest: for instance, POT is preferable for quantile estimation, while BM is preferable for return level estimation. Finally, we discuss the two approaches for multivariate observations and identify various open ends for future research.
\end{abstract}


\end{frontmatter}

\section{Introduction} \label{sec:intro}
Extreme-Value Statistics can be regarded as the art of extrapolation. Based on a finite sample from some distribution $F$,
typical quantities of interest are quantiles whose levels are larger than the largest observation or probabilities of rare events which have not occurred yet in the observed sample. Estimating such objects typically relies on the following fundamental domain-of-attraction condition: there exists a constant $\gamma\in\R$ and sequences
$a_r>0$ and $b_r$, $r\in\N$, such that
\begin{equation}\label{eq:doabm}
\lim_{r\to\infty} F^r(a_r x+b_r)=\exp\set{-(1+\gamma x)^{-1/\gamma}} \mbox{\ for all \ } 1+\gamma x>0.
\end{equation}
In that case, $\gamma$ is called the \textit{extreme value index}. The limit appears unnecessarily specific, but it is in fact the only non-degenerate limit of the expression on the left-hand side.  An equivalent representation of the domain of attraction condition \eqref{eq:doabm} is as follows: there exists a positive function $\sigma=\sigma(t)$ such that
\begin{equation}\label{eq:doapot}
\lim_{t\uparrow x^*} \frac{1-F(t+\sigma(t)x)}{1-F(t)}=(1+\gamma x)^{-1/\gamma} \mbox{\ for all \ } 1+\gamma x>0,
\end{equation}
where $x^*$ denotes the right end-point of the support of $F$, see \cite{BalDeh74}. The two sequences in \eqref{eq:doabm} are related to the function $\sigma$ as follows: $a_r=\sigma(b_r)$ and $b_r=U(r)$ where $U(r)=F^{\leftarrow}(1-1/r) = (1/(1-F))^{\leftarrow}(r)$, with $\cdot^{\leftarrow}$ denoting the left--continuous inverse of some monotone function.

Consider for instance the consequences of the previous two displays for high quantiles of $F$. By by \eqref{eq:doabm}, for all $p$ sufficiently small,
\begin{align} \label{eq:quantile}
F^{\leftarrow}(1-p) \approx b_r + a_r \frac{\{- r \log(1-p)\}^{-\gamma}-1}{\gamma} \approx b_r + a_r \frac{(rp)^{-\gamma}-1}{\gamma}.
\end{align}
Hence, by the plug-in-principle, a suitable choice of $r$ and suitable estimators of $a_r, b_r$ and $\gamma$ immediately suggest estimators for high quantiles.

Similarly, by \eqref{eq:doabm}, for all $p$ sufficiently small,
\begin{align} \label{eq:quantilepot}
F^{\leftarrow}(1-p) \approx t + \sigma(t) \frac{\{\frac{p}{1-F(t)}\}^{-\gamma}-1}{\gamma}.
\end{align}
Again, by the plug-in-principle, a suitable choice of $t$ and suitable estimators of $\sigma(t)$,  $\gamma$ and $1-F(t)$ immediately leads to estimators for high quantiles. Here, $t$ is typically  chosen as a large order statistic $t=X_{n-k:n}$ and $1-F(t)$ is replaced by $k/n$.

In practice, estimators for the parameters in these two approaches typically follow their corresponding basic principles: the block maxima method motivated by \eqref{eq:doabm} and the peak-over-threshold approach motivated by \eqref{eq:doapot}. Let $X_1, X_2, \dots, X_n$ be a sample of observations drawn from $F$, and for the moment assume that the observations are independent. Then \eqref{eq:doabm} gives rise to the \textit{block maxima method (BM)} \citep{Gum58}: for some block size $r\in\{1,\dots, n\}$, divide the data into $k=\lfloor n/r\rfloor$  blocks of length $r$ (and a possibly remaining block of smaller size which has to be discarded).
By independence, each block has cdf $F^r$. By \eqref{eq:doabm}, for large block sizes $r$,  the sample of block maxima can then be regarded as an approximate i.i.d.\ sample from the three-parametric generalized extreme-value (GEV) distribution $G_{\gamma, b,a}$  with location parameter $b=b_r$, scale parameter $a=a_r$ and shape parameter $\gamma$,  defined by its cdf
\[
G_{\gamma , b,a}^{\mathrm{GEV}}(x)
:= \exp\Big\{-\Big(1+\gamma \frac{x-b}{a}\Big)^{-1/\gamma}\Big\} \mathbf1\Big(1+\gamma \frac{x-b}{a}>0\Big).
\]
The three parameters can  be estimated by maximum-likelihood or moment-matching, among others. Irrespective of the particular estimation principle, any estimator defined in terms of the sample of block maxima will be referred to as an estimator based on the block maxima method.

Often, an available data-sample consists of block maxima only, for example, annual maxima of a river level. Then a practitioner may only rely on the block maxima method. If the underlying observations are available, then \eqref{eq:doapot} gives rise to the competing \textit{peak-over-threshold approach (POT)} \citep{Pic75}: for sufficiently large $t$ in \eqref{eq:doapot}, we obtain that, for any $x>0$,
\begin{align} \label{eq:POT}
\Pr(X>t + x\mid X>t) = \frac{\Pr(X>t + x)}{\Pr(X>t)} \approx \Big(1+\gamma \tfrac{x}{\sigma}\Big)^{-1/\gamma} =: 1- G_{\gamma , \sigma}^{\mathrm{GP}}(x),
\end{align}
where the right-hand side defines the two-parametric generalized Pareto (GP)  distribution with  scale parameter $\sigma:=\sigma(t)$ and shape parameter $\gamma$.
 In practice, $t$ is typically chosen as the $(n-k)$-th order statistic $X_{n-k:n}$ for some intermediate value $k$ (hence, $X_{n-k:n}$ is the $(1-1/r)$-sample quantile with $r=n/k$). Then, one may regard the sample $X_{n-k+1:n}-X_{n-k:n}, \dots X_{n:n}-X_{n-k:n}$  as observations from the two-parametric generalized Pareto-distribution. The parameters can hence be estimated by moment matching, and even by maximum-likelihood  since the sample of order statistics can actually be regarded as independent (see, e.g., Lemma~3.4.1 in \citealp{DehFer06}). In general, any estimator defined in terms of all observations exceeding some (random) threshold will be referred to as an estimator based on the POT approach. The vanilla estimator within this class is the Hill estimator \citep{Hil75} in the case $\gamma>0$.

The goal of the present paper is an in-depth comparison of the two approaches, in particular in terms of recent solid theoretical advances on asymptotic theory for the BM method, but also with a view on  time series data and multivariate observations. The discussion will mostly be of reviewing nature, but some new insights will be presented as well.
The next paragraphs summarize our contribution in a chronological order.

\smallskip

\begin{inparaenum}
\item \textbf{Efficiency comparison in i.i.d.\ scenarios.}
It seems to be general consensus among researchers in extreme value statistics that the POT method produces more efficient estimators than the BM method. The main heuristic reason is that all large observations are used for the calculation of POT estimators, while BM estimators may miss some large observations falling into the same block.
This heuristics was confirmed by simulation studies in \cite{Cai09}, see also  the additional references mentioned in \cite{FerDeh15}.
Due to some recent advances on theoretic properties of BM estimators \citep{Dom15, FerDeh15, BucSeg14, BucSeg18, DomFer17}, the two approaches may actually be compared on solid theoretical grounds.
For a certain type of cdfs, such a discussion has been carried out in \cite{FerDeh15} and \cite{DomFer17}; their findings are summarized and extended in Section~\ref{sec:iid} of this paper.
We show that, depending on the data generating process, the convergence rate of the two methods may be different, with no general winner being identifiable.
In case the rates are the same, BM estimators typically have a smaller variance, but a larger bias than their POT-competitors.

\smallskip

\item \textbf{BM and POT applied to time series.}
The above discussion motivating the BM and POT approach was based on an i.i.d.\ assumption on the underlying sample.
This assumption is actually quite restrictive since it excludes many common environmental or financial applications, where the underlying sample is typically a (stationary) time series. In this setting, it seems to be general consensus that the block maxima method still `works' because the block maxima are (1) still approximately GEV-distributed \citep{Lea74} and (2) distant from each other and thus bear low serial dependence. Consequently, the sample of block maxima may still be regarded as an approximate i.i.d.\ sample from the three-parametric GEV-distribution.
This heuristics is confirmed by recent theoretical results in \cite{BucSeg18, BucSeg14}.

Nevertheless, as discussed in Section~\ref{sec:ts} below, an obstacle occurs: the location and scale parameters attached to block maxima of a time series will typically be different from those of an i.i.d.\ series from the same stationary distribution $F$, whence estimators for quantities that depend on the stationary distribution only will possibly be inconsistent.
The missing link is provided by the extremal index \citep{Lea83}, a parameter in $[0,1]$ capturing the tendency of the extreme observations of a stationary time series to occur in clusters.
The discussion will be worked out on the example of high quantile estimation:
based on suitable estimators for the extremal index, see Section~\ref{sec:ts} below, \eqref{eq:quantile} can in fact be modified to obtain consistent BM estimators of large quantiles.

On the other hand, estimators based on the POT method for characteristics of the stationary distribution remain consistent. This however comes at the cost of an increased variance of the estimators due to potential clustering of extremes, see \citealp{Hsi91, Dre00, Roo09}, among many others.
Should the ultimate interest be in return level or return periods estimation, the picture is reversed: the BM method is consistent without the need of estimating the extremal index, while POT estimators typically require estimates of the extremal index.
More details are provided in Section~\ref{sec:ts}.

\smallskip

\item \textbf{Extensions to multivariate observations and stochastic processes.} The previous discussion focussed on the univariate case. Section~\ref{sec:mult} briefly discusses multivariate extensions. On the theoretical side, while there are many results available for the POT approach, there is clearly a supply issue regarding the BM approach: almost all statistical theory is  formulated under the assumption that the block maxima genuinely  follow a multivariate extreme value distribution, thereby ignoring a potential bias and rendering a fair theoretical comparison impossible for the moment (to the best of our knowledge, the only available results on the BM method are provided in \citealp{BucSeg14}). Instead, we provide a review on some of the existing theoretical results using these two approaches, and identify the open ends that may eventually lead to results allowing for an in-depth theoretical comparison in the future.

Not surprisingly, a fair comparison is even more difficult when considering extreme value analysis for stochastic processes. Most of the existing statistical methods are based on max-stable process models, i.e., on limit models arising for maxima taken over i.i.d.\ stochastic processes. The respective statistical theory is again mostly formulated under the assumption that the observations are genuine observations from the max-stable model, whence the statistical methods can (in most cases) be generically attributed to the BM approach. As for multivariate models, potential bias issues are mostly ignored. By contrast to multivariate models, however, very little is known for the POT approach to processes. A comparison is hence not feasible for the moment, and we limit ourselves to a brief review of existing results in Section~\ref{sec:process}.
\end{inparaenum}

\smallskip

Finally, we end the paper by a section summarizing possible open research questions, Section~\ref{sec:open}, and by a short conclusion, Section~\ref{sec:con}.

\section{Efficiency Comparison for univariate i.i.d.\ observations}
\label{sec:iid}
The efficiency of BM and POT estimators can be compared in terms of their asymptotic bias and variance. In this section, we particularly focus on the estimation of the extreme value index $\gamma$ because for estimating other tail related characteristics such as high quantiles or tail probabilities, the asymptotic distributions of  respective estimators are typically dominated by those derived from estimating the extreme value index.

In both the BM and POT approach, a key tuning parameter is the intermediate sequence $k=k(n)$, which corresponds to either the number of blocks in the BM approach, or the number of upper order statistics in the POT approach. For most data generating processes, consistency  of respective estimators can be guaranteed if $k$ is chosen in such a way that $k\to\infty$ and $k/n\to 0$ as $n\to\infty$. Here, the small fraction $k/n$ reflects the fact that the inference is based on observations in the tail only. Typically, the variance of respective estimators is of order $1/k$, while the bias depends on how well the distribution of block maxima or threshold exceedances is approximated by the GEV or GP distribution, respectively. Choosing $k$ in such a way that variance and squared bias are of the same order (see Section~\ref{subsec:rate} below), one may derive an optimal rate of convergence for a given estimator. Depending on the model, the optimal choice of $k$ may result in a faster rate for the BM method or the POT approach, as will be discussed next.

It is instructive to consider two extreme examples first (where the condition $k/n\to0$ as $n\to\infty$ may in fact be discarded): if $F$ is the standard Fr\'echet-distribution, then block maxima of size $r=1$ are already GEV-distributed. In other words a sample of $k=n$ block maxima of size $r=1$ can be used for estimation via the BM method. The rate of convergence is thus $1/\sqrt{n}$ and the POT method fails to achieve this rate. On the other hand, if $F$ is the standard Pareto distribution, then all $k=n$ largest order statistics can be used for the estimation via the POT approach. The rate of convergence is $1/\sqrt{n}$ for the POT method, which is not achievable via the BM method.

Apart form these two (or similar) extreme cases, the optimal choice of $k$ depends on second order conditions quantifying the speed of convergence in the domain of attraction condition. These are often (though not always) formulated in terms of the two quantile functions
\[
U(x) =  \Big(\frac{1}{1-F}\Big)^\leftarrow(x) \quad\text{and}\quad  V(x) = \Big(\frac{1}{-\log F}\Big)^\leftarrow(x)
\]
for the POT- and the BM method, respectively. Note that the domain of attraction condition~\eqref{eq:doabm} is equivalent to the fact that there exists a positive function $a_{\POT}$ such that, for all $x>0$,
\begin{align} \label{eq:doau}
\lim_{t \to \infty} \frac{U(tx)-U(t)}{a_{\POT}(t)} =  \int_1^x s^{\gamma-1}\, ds ,
\end{align}
see Theorems 1.1.6 and  1.2.1 in \cite{DehFer06}.
The function  $a_\POT$  is related to the sequence $(a_r)_r$ appearing in \eqref{eq:doabm} via $a_{\POT}(r) = a_{\lfloor r \rfloor}$.

In parallel, \eqref{eq:doabm} is also equivalent to the fact there exists a positive function $a_{\BM}$ such that, for all $x>0$,
\begin{align} \label{eq:doav}
\lim_{t \to \infty} \frac{V(tx)-V(t)}{a_{\BM}(t)} = \int_1^x s^{\gamma-1}\, ds.
\end{align}
The bias of certain BM- and POT estimators is determined by the speed of convergence in the latter two limit relations, which can be captured by suitable second order conditions.

For $\gamma\in\R,\rho\le 0$ and $x>0$, let
\[
h_\gamma(x) = \int_1^x s^{\gamma-1}\, ds, \qquad H_{\gamma,\rho}(x) = \int_1^x s^{\gamma-1} \int_1^s u^{\rho-1}\, du \, ds.
\]

\begin{definition}[Second order conditions] \label{def:secor}
Let $F$ be a cdf satisfying the domain-of-attraction condition \eqref{eq:doabm} for some $\gamma\in\R$. Consider the following two assumptions.
\begin{compactenum}[$(SO)_U$]
\renewcommand{\theenumi}{$\mathrm{(SO)_U}$}
\renewcommand{\labelenumi}{\theenumi}
\item\label{cond:sou}  Suppose that there exists $\rho_{\POT}\le 0$, a positive function $a_\POT$ and a positive or negative function $A_\POT$ with $\lim_{t\to\infty } A_\POT(t) = 0$, such that, for all $x>0$,
\[
\lim_{t \to \infty}  \frac{1}{A_\POT(t)} \bigg( \frac{U(tx)-U(t)}{a_{\POT}(t)} - h_\gamma(x) \bigg) = H_{\gamma, \rho_\POT}(x).
\]
\renewcommand{\theenumi}{$\mathrm{(SO)_V}$}
\renewcommand{\labelenumi}{\theenumi}
\item \label{cond:sov}
 Suppose that there exists $\rho_{\BM}\le 0$, a positive function $a_\BM$ and a positive or negative function $A_\BM$ with $\lim_{t\to\infty } A_\BM(t) = 0$, such that, for all $x>0$,
\[
\lim_{t \to \infty}  \frac{1}{A_\BM(t)} \bigg( \frac{V(tx)-V(t)}{a_{\BM}(t)} - h_\gamma(x) \bigg) = H_{\gamma, \rho_\BM}(x).
\]
\end{compactenum}
\end{definition}
The functions $|A_\BM|$ and $|A_\POT|$ are then necessarily regularly varying with index $\rho_\BM$ and $\rho_\POT$, respectively. The limit function $H_{\gamma,\rho}$ might appear unnecessarily specific, but in fact it is not, see \cite{DehSta96} or Section B.3 in \cite{DehFer06}. If the speed of convergence in \eqref{eq:doau} or \eqref{eq:doav} is faster than any power function, we set the respective second order parameter as minus infinity. For example, for $F=G_{\gamma, \sigma}^{\mathrm{GP}}$ from the GP family, we have $\{U(tx)-U(t)\}/(\sigma t^\gamma)=h_\gamma(x)$, i.e. $\rho_\POT = -\infty$ in this case.  Likewise, any $F=G_{\gamma,\sigma,\mu}^{\mathrm{GEV}}$ from the GEV distribution satisfies $\{V(tx)-V(t)\}/(\sigma t^\gamma)=h_\gamma(x)$, which prompts us to define  $\rho_\BM=-\infty$.

It is important to note that $\rho_{\BM}$ and $\rho_{\POT}$ can be vastly different. A general result can be found in \cite{DreDehLi03}, Corollary A.1: under an additional condition which only concerns the cases $\gamma=1$, $\rho_\BM=-1$ or $\rho_\POT=-1$, the two coefficients are equal within the range $[-1,0]$. Otherwise, if \ref{cond:sov} holds with $\rho_{\BM}<-1$, then \ref{cond:sou} holds with $\rho_{\POT}=-1$;  if \ref{cond:sou} holds with $\rho_{\POT}<-1$, then \ref{cond:sov} holds with $\rho_{\BM}=-1$. Some values of the parameters for various types of distributions are collected in Table~\ref{tab:secor}.

\begin{table}[t!]
\centering
\begin{tabular}{l |ccc}
  \hline
  \hline
  Distribution & $\gamma$ & $\rho_{\POT}$ & $\rho_{\BM}$   \\ \hline
  $\mathrm{GP}(\gamma,\sigma)$ & $\gamma$ & $- \infty$ & $-1$ \\
  $\mathrm{Exponential}(\lambda)$ & 0 & $ - \infty$ & $-1$ \\
  $\mathrm{Uniform}(0,1)$ & $-1$ & $-\infty$  & $-1$ \\
  $\mathrm{Arcsin}$ & $-2$ & $-2$ & $-1$ \\
  $\mathrm{Burr}(\eta,\tau,\lambda)$ & $1/(\lambda\tau) $ & $-1/\lambda$ & $\max(-1/\lambda, -1)$  \\
  $t_\nu, \nu\ne 1$ & $1/\nu $ & $-2/\nu$ & $\max(-2/\nu, -1)$  \\ \hline
  $\mathrm{Cauchy} (=t_1)$ & 1 & $-2$ & $-2$ \\
  $\mathrm{Weibull}(\lambda,\beta), \beta\ne 1$ & 0 & 0 & 0 \\
  $\Gamma(\alpha,\beta)$ & 0 & 0 & 0 \\
  $\mathrm{Normal}(\mu,\sigma^2)$ & 0 & 0 & 0 \\
  \hline
  $F(x) = \exp(-(1+x^\alpha)^\beta)$ & $1/(\alpha\beta)$ & $\max(-1/\beta, -1)$  & $-1/\beta$ \\
  $\mathrm{\text{Fr\'{e}chet}}(\alpha, \sigma)$ & $1/\alpha$ & $-1$ & $- \infty$ \\
  $\mathrm{\text{Reverse Weibull}}(\beta, \mu, \sigma)$ & $-1/\beta$ & $-1$ & $- \infty$ \\
  $\mathrm{GEV}(\gamma,\mu,\sigma)$ & $\gamma$ & $-1$ & $- \infty$ \\
    \hline \hline
\end{tabular}

\medskip
\caption{Extreme value index and second order parameters for various models. }\label{tab:secor}
\vspace{-.4cm}
\end{table}

We remark that for the the $t_1$-distribution, we obtained $\rho_{\BM}=\rho_{POT}=-2$. This is a special example for which Corollary 4.1 in \cite{DreDehLi03} is not applicable: $2tA(t)$ converges to $0=1-\gamma$. Notice that for the six models in the first category $\rho_{POT}<\rho_{BM}$ (if we consider $\lambda>1$ in the Burr distribution and $\nu>2$ in the $t_\nu$ distribution). For the four models in the second category $\rho_{POT}=\rho_{BM}$ while for the last four models, $\rho_{POT}>\rho_{BM}$ if we consider $\beta>1$ in the model $F(x) = \exp(-(1+x^\alpha)^\beta)$.

Let us now consider asymptotic theory for the estimation of the extreme value index~$\gamma$. Perhaps surprisingly, asymptotic theory for the BM method has hitherto mostly ignored the fact that block maxima are only approximately GEV distributed (see, e.g., \citealp{PreWal80, HosWalWoo85, BucSeg17}, among others). Only recent theoretical studies in \cite{FerDeh15} and \cite{DomFer17} for the probability weighted moment (PWM) and the maximum likelihood (ML) estimator, respectively, take the approximation into account. Correspondingly, the asymptotic bias can be explicitly analyzed, relying on the second order condition \ref{cond:sov} above. On the other hand, solid theoretical studies regarding the POT method have a much longer history, see \cite{DehFer06} for a comprehensive overview. For the sake of theoretical comparability with the BM method, we will subsequently exemplarily deal with the ML estimator and the PWM estimator, for which Theorems~3.4.2  and 3.6.1 in \cite{DehFer06} provide the respective  asymptotic theory under the assumption that \ref{cond:sou} is met (the results rely on \citealp{Dre98, DreDehFer04}).

Summarizing the above mentioned results, for both methods (BM and POT), the ML-estimators are  consistent for $\gamma>-1$ and asymptotically normal for $\gamma>-1/2$, while PWM-estimators are consistent for $\gamma<1$ and asymptotically normal for $\gamma<1/2$.  Asymptotic theory  is formulated under the conditions that $k=k_n$ satisfies $k\to\infty$ and $k/n\to0$ (POT method) or $r=r_n$ satisfies $r\to\infty$ and $k=r/n \to 0$ (BM method), as $n\to\infty$. Under the respective second order conditions \ref{cond:sou} and \ref{cond:sov} formulated above, the asymptotic results can be summarized as
\begin{align*}
\hat \gamma \stackrel{d}{\approx} \mathcal N \Big(\gamma + A_{m}(n/k) b , \frac{1}{k} \sigma^2\Big), \quad m \in \{\BM, \POT\},
\end{align*}
where $\hat\gamma$ is one of the four estimators, and where the asymptotic bias $b$ and the asymptotic variance $\sigma^2$ depend on the specific estimator, the second order index $\rho_m$ and $\gamma$.
In particular, the rate of convergence of the bias $A_m(n/k)$ crucially depends on the second order index $\rho_m$.

In the next two subsections, we first discuss the best achievable rate of convergence
and then the asymptotic mean squared error in case the rates are the same. Finally, in the last subsection, we discuss the choice of $k$, i.e., the number of large order statistics in the POT approach or the number of blocks in the BM approach.

\subsection{Rate of convergence} \label{subsec:rate}
As is commonly done, we consider the rate of convergence of the root mean squared error. It is instructive to first elaborate on the case $A_m(t) \asymp t^{\rho_m}$ with $\rho_m\in(-\infty,0)$. The best attainable rate of convergence is achieved when squared bias and variance are of the same order, that is, when
\[
A_m^2\Big(\frac{n}k\Big) \asymp \Big(\frac{n}k\Big)^{2\rho_m} \asymp \frac1k.
\]
Solving for $k$ yields $k \asymp n^{-2\rho_m/(1-2\rho_m)}$, which implies
\[
\text{Rate of Convergence of $\hat \gamma$} = n^{\rho_m/(1-2\rho_m)}
\]
irrespective of $m\in\{\ML, \PWM\}$. For the POT approach, this result is known to hold for many  other estimators of $\gamma$; see \cite{DehFer06} (though not for every estimator, see Table 3.1 in that reference). In fact, it can be shown that this is the optimal rate under some specific assumptions on the data generating process, see \cite{HalWel84}. We conjecture that the same result holds true for many other estimators relying on the BM method, though the literature does not provide sufficient theoretical results yet except for the ML and PWM estimators.

Since $\rho_{\BM}$ and $\rho_{\POT}$ might not be the same, the best attainable rate of convergence may be different for the BM and POT approach. Table~\ref{tab:mse1} provides a summary of which method results in a better rate. The case where the rates are the same is discussed in more detail in Section~\ref{subsec:bias} below.

\begin{table}[h!]
\centering
\begin{tabular}{c|ccc}
  \hline
  \hline
  2nd Order Parameters & Rate POT & Rate BM & Better rate \\ \hline
  $\rho=\rho_\BM=\rho_{\POT} \in [ -1, 0)$ & $n^{\rho/(1-2\rho)}$  &  $n^{\rho/(1-2\rho)}$ &   - \\
  $\rho_\BM= -1, \rho_\POT < -1$ & $n^{\rho_\POT/(1-2\rho_\POT)}$  & $n^{-1/3}$ & POT \\
  $\rho_\POT = -1, \rho_\BM< -1$ & $n^{-1/3}$ & $n^{\rho_\BM/(1-2\rho_\BM)}$ & BM \\
  \hline \hline
\end{tabular}
\medskip
\caption{Best attainable convergence rates for the BM and POT approach in case $A_m(t) \asymp t^{\rho_m}$ with $\rho_m<0$ and for typical relationships between $\rho_\BM$ and $\rho_\POT$ (see \citealp{DreDehLi03}).}\label{tab:mse1}
\end{table}

Let us finally mention that the specific assumption on the function $A_m$ made above (i.e., $A_m(t) \asymp t^{\rho_m}$ with $\rho_m\in(-\infty,0)$) is not essential, see the argumentation on pages~79--80 in \cite{DehFer06}. Moreover,  for $\rho_m=-\infty$, the convergence rate is
`faster than $n^{-1/2+\eps}$ for any $\eps>0$', and, depending on the underlying distribution, in fact could even achieve $n^{-1/2}$ (see also Remark~3.2.6 in \citealp{DehFer06}).

\subsection{Asymptotic mean squared error} \label{subsec:bias}
As discussed in the previous subsection, if $\rho_{\POT}\neq \rho_{\BM}$, the approach corresponding to a lower $\rho$ generically yields estimators for  $\gamma$ with a faster attainable rate of convergence than the other approach. In this subsection, we consider the case $\rho_{\POT}=\rho_{\BM}$. Then both approaches, at their best attainable rate of convergence, will yield estimators of $\gamma$ with the same speed of convergence. Hence, the efficiency comparison should be made at the level of asymptotic mean squared error (AMSE) or, more precisely,  its two subcomponents: asymptotic bias and asymptotic variance. Notice that the asymptotic bias and variance depends on the specific estimator used, whence the comparison can only be performed based on some preselected estimators.

A detailed analysis of the PWM and the ML estimators under the BM and POT approach has been carried out in \cite{FerDeh15} and \cite{DomFer17}, for the case $\rho_\BM = \rho_\POT \in [-1,0]$ and $\gamma\in(-0.5, 0.5)$. The results are as follows: when using the same value for $k$, being either the number of large order statistics in the POT approach or the number of blocks in the BM approach, the BM version of either ML or PWM leads to a  lower asymptotic variance compared to the corresponding POT version, for all $\gamma\in(-0.5, 0.5)$. On the other hand, the (absolute) asymptotic bias is smaller for the POT versions of the two estimators, for all $(\gamma, \rho) \in (-0.5, 0.5) \times [-1,0]$.

When comparing the optimal AMSE (where optimal refers to the fact that the parameter $k$ is chosen in such a way that the AMSE for the specific estimator is minimized), it turns out that, for the ML estimator, the POT approach yields a smaller optimal AMSE. For the PWM estimator, the BM method is preferable for most combinations of $(\gamma, \rho)$. When comparing all four estimators, the combination ML-POT has the overall smallest optimal AMSE.

\subsection{Threshold and block length choice}

Both the POT and the BM approach require a practical selection for the intermediate sequence $k=k_n$  in a sample of size~$n$. In the POT approach, the choice of $k$ problem can be interpreted as the choice of the threshold above which the POT approximation in \eqref{eq:POT} is regarded as sufficiently accurate. Similarly, in the BM approach, $k$ is related to $r=n/k$, which is the size of the block of which the GEV approximation to the block maximum is regarded as sufficiently accurate.

The theoretical conditions that $k\to\infty$ and $k/n\to 0$, as $n\to\infty$ are useless in guiding the practical choice. Practically, often a plot between the estimates based on various $k$ against the values of $k$ is made for resolving this problem, the so-called ``Hill plot'' \citep{DreDehRes00}, despite the fact that it can be also be applied to other POT or even BM estimators than just the Hill estimator. The ultimate choice is then made by taking a $k$ from the first stable region in the ``Hill plot''. Nevertheless, the estimators are often rather sensitive to the choice of $k$.

For the POT approach, there exist a few attempts on resolving the choice of $k$ issue in a formal manner. For example, one solution is to find the optimal $k$ that minimizes the asymptotic MSE; see, e.g., \cite{DanDehPenDev01}, \cite{DreKau98} and \cite{GuiHal01}. As an indirect solution to the problem, one may also rely on bias corrections, which typically allows for a much larger choice of $k$, see, e.g., \cite{GomDehRod08}. After the bias correction, the ``Hill plot'' usually shows a stable behavior and the estimates are less sensitive to the choice of $k$. For an extensive review on bias corrections, see \cite{BeiCaeGom12}.

Compared to the extensive studies on the threshold choice and on bias corrections for the POT approach, there is, to the best of our knowledge, no existing literature addressing these issues for the BM approach. This may partly be explained by the fact that block sizes are often given by the problem at hand, for instance, block sizes corresponding to year. Nevertheless, based on the recent solid theoretical advances on the BM method, the foundations are laid to explore these issues in a rigorous manner in the future.

\section{BM and POT for Univariate Stationary Time Series} \label{sec:ts}

In many practical applications, the discussion from the previous section is not quite helpful: the underlying data sample is not i.i.d., but in fact a stretch of a possibly non-stationary time series. Often, by either restricting attention to a proper time horizon or by some suitable transformation, the time series can at least be assumed to be stationary.\footnote{For example, for financial applications, the stationarity assumption can often be approximately guaranteed by restricting attention to a time horizon during which few macro economic conditions had changed. Similarly, for environmental applications, this can be achieved by restricting attention to observations falling into, say,  the summer months.}  Throughout this section, we make the following generic assumption: $(X_t)_{t\in\Z}$ is a strictly stationary univariate time series, and the stationary cdf $F$ satisfies the domain-of-attraction condition~\eqref{eq:doabm}.
It is important to note that the parameters $\gamma,a_r$ and $b_r$ only depend on the stationary cdf $F$, and that for instance \eqref{eq:quantile} expressing high quantiles of $F$ through these parameters continues to hold for time series. Let us begin by passing over the arguments from Section~\ref{sec:intro} that eventually led to the BM- and POT method.

\subsection{The POT approach for time series}
Recall that the POT approach is based on the sample of large order statistics denoted by $\mathcal{X}_\POT= \{X_{n-k:n}, \dots, X_{n:n}\}$. The main motivation that lead us to consider this sample was the marginal limit relation~\eqref{eq:doapot}. Bearing in mind that, under mild extra conditions  on the serial dependence (ergodicity, mixing conditions, \ldots), empirical moments are consistent for their theoretical counterparts, it is thus still reasonable to estimate the respective parameters by any form of moment matching, e.g., by PWM.
The asymptotic variance of such estimators will however be different from the i.i.d.\ case in general (a consequence of central limit theorems for time series under mixing conditions).

Consider the ML-method:  unlike for i.i.d.\ data, the sample $\mathcal{X}_\POT$ cannot be regarded as independent anymore, whence it is in general impossible to derive the (approximate) generalized Pareto likelihood of $\mathcal{X}_\POT$. As a circumvent, one may `do as if' the likelihood arising in the i.i.d.\ case is also the likelihood for the time series case (quasi-maximum likelihood), and use essentially the same ML-estimators as for the i.i.d.\ case. Then, since the latter estimator is in fact also depending on empirical moments only, we still obtain proper asymptotic properties such as consistency and asymptotic normality.

Respective theory can be found in \cite{Hsi91, ResSta98} for the Hill estimator and in \cite{Dre00} for a large class of estimators, including PWM and ML. Most of the estimators have the same bias as in the i.i.d.\ case, whereas their asymptotic variances depend on the serial dependence structure and are usually higher than those  obtained in the i.i.d.\ case. Since the asymptotic bias shares the same explicit form, bias correction can also be performed in the same way as in the i.i.d.\ case; see, e.g., \cite{DehMerZho16}.

\subsection{The BM approach for time series}
Recall that the BM approach is based on the sample of block maxima $\mathcal{X}_\BM= \{M_{1,r}, \dots, M_{k:r}\}$, where $M_{j,r}$ denotes the maximum within the $j$th disjoint block of observations of size $r$. The main motivation in Section~\ref{sec:intro} that lead us to consider this sample as approximately GEV-distributed was the relation
\[
\Pr(M_{1,r}\le a_rx+b_r) = F^r(a_rx+b_r) \approx G_{\gamma,0,1}^\GEV(x),
\]
for large $r$. The first equality is not true for time series, whence more sophisticated arguments must be found for the BM method to work for time series.
In fact, it can be shown that if $F$ satisfies \eqref{eq:doabm}, if $\Pr(M_{1,r}\le a_rx+b_r)$ is convergent for some $x$ and if mild mixing conditions on the serial dependence (known as $D(u_n)$-conditions) are met, then there exists a constant $\theta\in[0,1]$ such that
\[
\lim_{r\to\infty} \Pr(M_{1:r} \le a_r x + b_r ) = \big( G_{\gamma,0,1}^\GEV(x) \big)^\theta
\]
for all $x\in\R$ \citep{Lea83}. The constant $\theta$ is called the \textit{extremal index} and can be interpreted as capturing  the tendency of the time series that extremal observations occur in clusters. If $\theta>0$, then letting
\begin{align} \label{eq:tilde}
\tilde a_r = a_r \theta^\gamma, \qquad \tilde b_r = b_r - a_r \frac{1-\theta^\gamma}{\gamma}
\end{align}
we immediately obtain that
\begin{align} \label{eq:doats}
\lim_{r\to\infty} \Pr(M_{1:r} \le \tilde a_r x + \tilde b_r ) = G_{\gamma,0,1}^\GEV(x)
\end{align}
for all $x\in\R$. Hence, the sample $\mathcal{X}_\BM$ is approximately GEV-distributed with parameter $(\tilde a_r, \tilde b_r, \gamma)$, which can then be estimated by any method of choice. It is important to note that, unless $\theta=1$ or $\gamma=0$,  $\tilde a_r $ and $\tilde b_r$ are different from $a_r$ and $b_r$. Consequently, additional steps must be taken for estimating quantiles of $F$ via \eqref{eq:quantile}, see also Section~\ref{subsec:quantile} below. Via \eqref{eq:tilde}, it is possible to transform between $(a_r,b_r)$ and $(\tilde a_r, \tilde b_r)$ if the extremal index $\theta$ is known or estimated. Regarding the estimation of the extremal index, a large variety of estimators has been proposed, which may itself be grouped into four categories: 1) BM-like estimators based on ``blocking'' techniques (\citealp{, Nor15, BerBuc17}), 2) POT-like estimators that rely on threshold exceedances (\citealp{FerSeg03, Suv07}), 3) estimators that use both principles simultaneously (\citealp{Hsi93, Rob09, RobSegFer09}) and 4) estimators which, next to choosing a threshold sequence, require the choice of a \textit{run-length} parameter \citep{SmiWei94, WeiNov98}.

Since the distance between the time points at which the maxima within two successive blocks are attained is likely to be quite large, the sample $\mathcal X_\BM$ can be regarded as approximately independent. As a matter of fact, the literature on statistical theory for the BM method is mostly based on the assumption that $\mathcal X_\BM$ is a genuine i.i.d.\ sample from the GEV-family (see, e.g., \citealp{PreWal80, HosWalWoo85, BucSeg17}, among others). Two approximation errors are thereby completely ignored: the cdf is only approximately GEV, and the sample is only approximately independent. Solid theoretical results taking  these errors into account are rare: \cite{BucSeg18} treat the ML-estimator in the heavy-tailed case ($\gamma>0$). The main conclusions are: the sample can safely be regarded as independent, but a bias term may appear which, similar as in Section~\ref{sec:iid}, depends on the speed of convergence in \eqref{eq:doats}. \cite{BucSeg18b} improve upon that estimator by using sliding blocks instead of disjoint blocks. The asymptotic variance of the estimator decreases, while the bias stays the same. Moreover, the resulting `Hill-Plots' are much smoother, guiding a simpler choice for the block length parameter.

\subsection{Comparison between the two methods} \label{subsec:compts}
Let us summarize the main conceptual differences between the BM and the POT method for time series. First of all, BM and POT estimate `the same' extreme value index $\gamma$, but possibly different scaling sequence $\tilde a_r, \tilde b_r$ and $a_r, b_r$. Second,  the sample $\mathcal X_\BM$ can be regarded as asymptotically independent (asymptotic variances of estimators are the same as if the sample was i.i.d.), while $\mathcal X_\POT$  is serially dependent, possibly increasing asymptotic variances of estimators compared to the i.i.d. case.

Due to the lack of a general theoretical result on the BM method, a theoretical comparison on which method is more efficient  along the lines of Section~\ref{sec:iid} seems out of reach for the moment.  In particular, a relationship between the respective second order conditions controlling the bias is yet to be found. However, some insight into the merits and pitfalls of two approaches can be gained by considering the problem of estimating high quantiles and return levels.

\subsubsection{Estimating high quantiles} \label{subsec:quantile}
Recall that high quantiles of the stationary distribution can be expressed in terms of $a_r, b_r$ and $\gamma$, see \eqref{eq:quantile}. As a consequence, based on the plug-in principle, the POT method immediately yields estimators for high quantiles. On the other hand, the BM method cannot be used straight-forwardly, as it commonly only provides estimators of $\tilde a_r, \tilde b_r$ and $\gamma$. Via \eqref{eq:tilde}, the latter estimators may be transferred into estimators of $a_r, b_r$ and $\gamma$ using an additional estimator of the extremal index~$\theta$. It is important to note that the latter estimators typically depend on the choice of one or two additional parameters, and that they are often quite variable. By contrast, the POT approach therefore seems more suitable when estimating high quantiles or, more generally, parameters that only depend on the stationary distribution (such as probabilities of rare events). Recall though that  estimators based on the POT approach usually suffer from a higher asymptotic variance due to the serial dependence.

\subsubsection{Estimating return levels} \label{subsec:quantile}
Let $F_r(x)=\Pr(M_{1:r} \le x)$. For $T \ge 1$, the $T$-return level of the sequence of  block maxima is defined as the $1-1/T$ quantile of $F_r$,
that is,
\[
  \mathrm{RL}(T,r) = F_r^{\leftarrow}(1-1/T) = \inf \{ x \in \R : F_r(x) \ge 1 - 1/T \}.
\]
Since block maxima are asymptotically independent, it will take on average $T$ blocks of size $r$ until the first such block whose maximum exceeds $\mathrm{RL}(T, r)$.
Now, since $F_r$ is approximately equal to the GEV-cdf with parameters $\gamma, \tilde b_r, \tilde a_r$ for large $r$ by \eqref{eq:doats}, we obtain that
\[
  \mathrm{RL}(T,r) \approx \tilde b_r + \tilde a_r \frac{\{- r \log(1-1/T)\}^{-\gamma}-1}{\gamma} \approx \tilde b_r + \tilde a_r \frac{(r/T)^{-\gamma}-1}{\gamma}.
\]
In comparison to the estimation of high-quantiles, see \eqref{eq:quantile}, we have now expressed the object of interest in terms of the sequences $\tilde a_r$ and $\tilde b_r$ and the extreme-value index $\gamma$.  Following the discussion in the previous section, it is now the BM method which yields simpler estimators that do not require additional estimation of the extremal index. By contrast, the POT approach only results in estimators of $(a_r,b_r)$ and $\gamma$, and therefore requires a transformation to $(\tilde a_r,\tilde b_r)$ via \eqref{eq:tilde} based on an estimate of the extremal index~$\theta$.

\section{BM and POT for Multivariate Observations}  \label{sec:mult}

Due to the lack of asymptotic results on the multivariate BM method which take the approximation error into account, a deep comparison between the BM and POT approach is not feasible yet. Within this section we try to identify the open ends that may eventually lead to such results in the future.

Let $F$ be a $d$-dimensional cdf. The basic assumption of multivariate extreme-value theory, generalizing \eqref{eq:doabm}, is as follows: suppose that there exists a non-degenerate cdf $G$ and sequences $(a_{r,j})_{r\in\N}, (b_{r,j})_{r\in\N}, j=1, \dots d,$ with $a_{r,j}>0$ such that
\begin{align} \label{eq:doamult}
\lim_{r \to\infty} \Pr\Big(\frac{\max_{i=1}^r X_{i,1}-b_{r,1}}{a_{r,1}} \le x_1  , \dots,\frac{\max_{i=1}^r X_{i,d}-b_{r,d}}{a_{r,d}}  \le x_d\Big)
= G(x_1, \dots, x_d)
\end{align}
for any $x_1, \dots, x_d\in\R$, where $\bm X_i=(X_{i,1}, \dots, X_{i,d})', i\in\N,$ is an i.i.d.\ sequence from $F$, and where the marginal distributions $G_j$ of $G$, $j=1, \dots, d$, are GEV-distributions with location parameter 0, scale parameter 1 and shape parameter $\gamma_j\in\R$ (location 0 and scale 1 can always be reached by adapting the sequences  $a_{r,j}$ are $b_{r,j}$  if necessary). The dependence between the coordinates of $G$ can be described in various equivalent ways (see, e.g., \citealp{Res87, BeiGoeSegTeu04, DehFer06}): by the stable tail dependence function $L$ \citep{Hua92}, by the exponent measure $\mu$ \citep{BalRes77},  by the Pickands dependence function $A$ \citep{Pic81}, by the tail copula $\Lambda$ \citep{SchSta06}, by the spectral measure $\Phi$ \citep{DehRes77}, by the madogram $\nu$ \citep{NavGuiCooDie09}, or by other less popular objects. All these objects are in one-to-one correspondence, and for each of them a large variety of estimators has been proposed, both in a nonparametric way and under the assumption that the objects are parametrized by an Euclidean parameter.

In this paper, we will mainly focus on nonparametric estimation. As in the univariate case, the estimators may again be grouped into BM and POT based estimators, see Sections~\ref{subsec:potmult} and \ref{subsec:bmmult} below. Often, estimation of the marginal parameters and of the dependence structure is treated successively. It is important to note that standard errors for estimators of the dependence structure may then be influenced by standard errors for the marginal estimation, a point which is often ignored in the literature on statistics for multivariate extremes. In fact, a phenomenon well-known in statistics for copulas \citep{GenSeg10} may show up: possibly completely ignoring additional information about the marginal cdfs, estimators for the dependence structure may have a lower asymptotic variance if marginal cdfs are estimated nonparametrically; see \cite{Buc14} for a discussion of the empirical stable tail dependence function from Section~\ref{subsec:potmult} below, and \cite{GenSeg09} for estimation of Pickands dependence function based on i.i.d.\ data from a bivariate extreme value distribution, Section~\ref{subsec:bmmult} below.

\subsection{The POT method in the multivariate case}  \label{subsec:potmult}

Suppose $\bm X_1, \dots, \bm X_n$, with $\bm X_i=(X_{i,1}, \dots, X_{i,d})'$, is an i.i.d.\ sample from $F$. Recall that the univariate POT method was based on the observations $\mathcal X_\POT=\{X_{n-k:n}, \dots, X_{n:n}\}$, which may be rewritten as $\mathcal X_\POT=\{X_i: \text{rank}(X_i \text{ among } X_1, \dots, X_n) \ge n-k) $. Thus, a possible generalization to multivariate observations consists of defining
\[
\mathcal X_\POT = \{ \bm X_i \mid
\text{rank}(X_{i,j} \text{ among } X_{1,j}, \dots, X_{n,j}) \ge n-k  \text{ for some } j=1, \dots, d\},
\]
that is, $\mathcal X_\POT$ comprises all observations for which at least one coordinate is large. Any estimator defined in terms of these observations may be called an estimator based on the multivariate POT method.

As an example, consider the estimation of the so-called stable tail dependence function~$L$, which is defined as
\begin{align} \label{eq:L}
L(\bm x) = \lim_{t\downarrow0} t^{-1} \Pr( F_1(X_1) > 1-tx_1 \text{ or } \dots \text{ or } F_d(X_d) > 1-tx_d),
\end{align}
where $\bm x=(x_1, \dots, x_d)'\in[0,1]^d$; a limit that necessarily exists under \eqref{eq:doamult}, but may also exist for marginals $F_j$ not in any domain-of-attraction. The function $L$ can be estimated by its empirical counterpart, defined as
\[
\hat L(x_1, \dots, x_d) =  \frac1k \sum_{i=1}^n \bm 1( \hat F_{n,1} (X_{i,1}) > 1-\tfrac{k}n x_1 \text{ or } \dots \text{ or } \hat F_{n,d} (X_{i,d}) > 1-\tfrac{k}n x_d),
\]
where $\hat F_{n,j}$ denotes the empirical cdf based on the observations $X_{1,j}, \dots , X_{n,j}$; see, e.g., \cite{Hua92}.  Since $\bm x\in [0,1]^d$, the estimator in fact only depends on the sample $\mathcal X_\POT$.

Suppose the following natural second order condition quantifying the speed of convergence in \eqref{eq:L} is met: there exists a positive or negative function $A$ and a real-valued function $g\not \equiv 0$ such that
\begin{align} \label{eq:secorm}
\lim_{t\to\infty} \frac{t \Pr( F_1(X_1) > 1-\tfrac{x_1}t \text{ or } \dots \text{ or } F_d(X_d) > 1-\tfrac{x_d}t) - L(x_1, \dots, x_d)}{A(t)} = g(\bm x)
\end{align}
uniformly in $\bm x\in[0,1]^d$. Then, under additional smoothness conditions on $L$, it can be shown that $\hat L$ is consistent and asymptotically Gaussian in terms of functional weak convergence, the variance being of order $1/k$ and the bias being of order $A(n/k)$, provided that  $k=k_n\to\infty$ and $k/n\to0$ as $n\to\infty$; see, e.g., \cite{Hua92, EinKraSeg12}, among others. Following the discussion in Section~\ref{sec:iid}, if we additionally assume that $A(t) \asymp t^\rho$ for some $\rho\in(-\infty,0)$, the best attainable convergence rate, achieved when squared bias and variance are balanced, is
\[
\text{Rate of Convergence of $\hat L(\bm x)$} = n^{\rho/(1-2\rho)}.
\]
This convergence rate is in fact optimal under additional conditions on the data-gene\-rating process, see \cite{DreHua98}. Also note that $\hat L$ suffers from an asymptotic bias as in the univariate case, and that corresponding bias corrections for the bivariate case have been proposed in \cite{FouDehMer2015}.

As in the univariate case, the literature on further theoretical foundations for the multivariate POT method is vast, see, e.g., \cite{EinDehPit01, EinSeg09} for nonparametric estimation of the spectral measure, \cite{DreDeh15} for estimation of failure probabilities, or  \cite{DehNevPen08, EinKraSeg12} for parametric estimators, among many others.

\subsection{The BM method in the multivariate case} \label{subsec:bmmult}
Again suppose $\bm X_1, \dots, \bm X_n$ is an i.i.d.\ sample from $F$. Let $r$ denote a block size, and $k=\lfloor n/r \rfloor$ the number of blocks.
For $\ell=1, \dots, k$, let $\bm M_{\ell,r} =(M_{\ell, 1,r}, \dots, M_{\ell,1,r})'$ denote the vector of componentwise block-maxima in the $\ell$th block of observations of size $r$ (it is worthwhile to note that $\bm M_{\ell,r}$ may be different from any $\bm X_{i}$). Any estimator defined in terms of the sample $\mathcal X_{\BM} = (\bm M_{1,r}, \dots, \bm M_{k,r})$ is  called an estimator based on the BM approach.

Just as for the univariate BM method, asymptotic theory is usually formulated under the assumption that $\bm M_1, \dots, \bm M_k$ is a genuine i.i.d.\ sample from the limiting distribution $G$; a potential bias is completely ignored. Moreover, estimation of the marginal parameters is often disentangled from estimation of the dependence structure, with theory for the latter either developed under the assumption that marginals are completely known (which usually leads to wrong asymptotic variances), or under the assumption that marginals are estimated nonparametrically. See, for instance, \cite{Pic81, CapFouGen97, ZhaWelPen08, GenSeg09, GudSeg12} for nonparametric estimators and \cite{GenGhoRiv95, DomEngOes16b} for parametric ones, among many others.

To the best of our knowledge, the only reference that takes the approximation error induced by the assumption of observing data from a genuine extreme-value model into account is \cite{BucSeg14}, where the estimation of the Pickands dependence function $A$ based on the BM-method is considered. Not only the bias is treated carefully there, but also the underlying observations $\bm X_1, \dots, \bm X_n$ may possess serial dependence in form of a stationary time series.  Just like in the univariate case described above, the best attainable convergence rate of the estimator again depends on a second order condition.

\subsection{Comparison between the two methods}
Due to the lack of honest theoretical results on the BM method, not much can be said yet about which method is better in terms of, say, the rate of convergence.  The missing tool is a multivariate version of Corollary A.1 in \cite{DreDehLi03}, allowing one to move from a BM second order condition (such as the one imposed in \citealp{BucSeg14}) to a POT second order condition as in \eqref{eq:secorm}, and vice versa. It then seems likely that similar phenomena as in the univariate case in Section~\ref{sec:iid} may show up.

\subsection{Multivariate time series}
Moving from i.i.d.\ multivariate observations to multivariate strictly stationary time series induces similar phenomena as in the univariate case, whence we keep the discussion quite short. Under suitable conditions on the serial dependence, estimators based on the POT approach are still consistent and asymptotically normal, though with a possibly  different asymptotic variance (this can for instance be deduced from \citealp{DreRoo10}). Regarding the BM method, the same heuristics as in the univariate case apply: block maxima may safely be assumed as independent and as following a multivariate extreme value distribution \citep{BucSeg14}. The estimators based on the BM method are then also consistent and asymptotically normal with a potential bias. Similar to the discussion on the location and scale parameters in the univariate case, the objects that are estimated by POT and BM may be different but are linked by the multivariate extremal index (\citealp{Nan94}, see also Section 10.5 in \citealp{BeiGoeSegTeu04}). Hence, following the discussion in Section~\ref{subsec:compts}, it seems  preferable to estimate quantities that only depend on the tail of the stationary distribution by the POT approach, while tail quantities similar to the univariate return levels (that also depend on the serial dependence) are preferably estimated by the BM approach. As in the univariate case, a detailed theoretical comparison does not seem to be feasible.

\section{BM and POT for stochastic processes} \label{sec:process}
The BM approach for stochastic processes is based on modeling by max-stable processes, i.e., on limit models arising for block maxima taken over i.i.d.\ stochastic processes.  Recent research has  focussed on the structure and characteristics of max-stable processes, see, e.g., \cite{Deh84}, \cite{GinHahVat90} and \cite{KabSchDeh09}; on simulating from max-stable processes, see, e.g., \cite{DomEyiRib13},  \cite{DieMik15}, \cite{DomEngOes16} and \cite{OesSchZhou18}; and on statistical inference based on max-stable processes, see, e.g., \cite{ColTaw96}, \cite{BuiDehZho08}, \cite{PadRibSis10} and \cite{HusDav14}.

As mentioned in the introduction, there is a clear supply issue regarding the POT approach to stochastic process models. Early studies such as \cite{EinLin06} consider the estimation of marginal parameters only, or consider nonparametric estimation of the dependence structure \citep{deHaanLin2003}, however with only weak consistency established. Recent development on \textit{Generalized Pareto Processes} allow for considering parametric estimation for the dependence structure, see, e.g. \cite{FerDeh14}, \cite{ThiOpi15} and \cite{HusWad17}. Given the imbalanced nature, we skip a deeper review on the BM and POT approaches for extremes regarding stochastic processes.

\section{Open problems}
\label{sec:open}

Throughout this paper, we have already identified a number of open research problems, mostly related to an honest verification of the BM approach. Within the following list, we recapitulate those issues and add several further possible research questions:

\medskip

 \begin{compactitem}

 \item Asymptotic theory on further estimators based on the block maxima method, if possible allowing for a comparison between the imposed second order condition and those from the POT approach.

 \item In case the BM method yields to faster attainable rates of convergence than the POT approach (Section~\ref{subsec:rate}): are the obtained rates optimal?

 \item Derive a test for which approach is preferably for a given data set ($H_0: \rho_\BM \le \rho_\POT$, or similar).

 \item Block length choice and bias reduction for BM.

 \item More results on the sliding block maxima method (non-heavy tailed case, multivariate case).

 \item A comparison of BM and POT second order conditions in the multivariate case.

 \item A comparison of return level/quantile estimation based on BM and POT, possibly incorporating an estimator for the extremal index.

 \item Extension to stochastic processes (max-stable processes and generalized Pareto processes): theoretical results on statistical methodology are still rare, and a comparison between BM and POT is not feasible yet.

 \end{compactitem}

\section{Conclusion}
\label{sec:con}

 There is no winner.

\bibliographystyle{imsart-nameyear}
\bibliography{biblio}

\end{document}